\documentclass[twocolumn,prd,nofootinbib,aps,prl,floats,floatfix,amsmath,amssymb,longbibliography,secnumarabic]{revtex4-2} 
\usepackage[final]{graphicx}
\usepackage{hyperref}
\usepackage{amsmath}
\usepackage{bbm}
\usepackage{amsfonts}
\usepackage{amssymb}
\usepackage{latexsym}
\usepackage{graphicx}
\usepackage[english]{babel}
\usepackage{multirow}
\usepackage{float}
\usepackage{url}
\usepackage{slashed}
\usepackage{xcolor} 
\usepackage[utf8]{inputenc}
\usepackage{verbatim}
\usepackage{stackengine}

\def\sfrac#1#2{{\textstyle{#1\over #2}}}
\newcommand{\be}{\begin{equation}}
\newcommand{\ee}{\end{equation}}
\newcommand{\ba}{\begin{array}}
\newcommand{\ea}{\end{array}}
\newcommand{\bea}{\begin{eqnarray}}
\newcommand{\eea}{\end{eqnarray}}
\newcommand{\sss}{\scriptscriptstyle}

\newcommand{\nn}{\nonumber}
\renewcommand{\L}{{\sss L}}

\newcommand{\W}{{\sss W}}
\newcommand{\Z}{{\sss Z}}
\newcommand{\F}{{\sss F}}

\newcommand{\diff}{\mathrm{{d}}}
\newcommand{\half}{\frac{1}{2}}

\begin{document}

\title{Sterile neutrino dark matter catalyzed by a very light dark photon
}

\author{Gonzalo Alonso-\'Alvarez}
\email{galonso@physics.mcgill.ca}
\thanks{ORCID: \href{https://orcid.org/0000-0002-5206-1177}{0000-0002-5206-1177}}
\affiliation{McGill University, Department of Physics, 3600 University St.,
Montr\'eal, QC H3A2T8 Canada}
\author{James M.\ Cline}
\email{jcline@phhysics.mcgill.ca}
\thanks{ORCID: \href{https://orcid.org/0000-0001-7437-4193}{0000-0001-7437-4193}}
\affiliation{McGill University, Department of Physics, 3600 University St.,
Montr\'eal, QC H3A2T8 Canada}
\begin{abstract}
Sterile neutrinos ($\nu_s$) that mix with active neutrinos ($\nu_a$) are interesting dark matter candidates with a rich cosmological and astrophysical phenomenology.
In their simplest incarnation, their production is severely constrained by a combination of structure formation observations and X-ray searches.
We show that if active neutrinos couple to an oscillating condensate of a very light $L_{\mu}-L_{\tau}$ gauge field, resonant $\nu_a$-$\nu_s$ oscillations can occur in the early universe, consistent with $\nu_s$ 
constituting all of the dark matter, while respecting X-ray constraints on
$\nu_s\to\nu_a\gamma$ decays.  Interesting deviations from standard solar and atmospheric neutrino oscillations can persist to the present.

\end{abstract}
\maketitle

\section{Introduction}
The possibility that sterile neutrinos $\nu_s$ constitute the dark matter (DM) of the universe has attracted steady interest since its inception long ago.  Supposing that the $\nu_s$ abundance is initially negligible, its relic density can be generated by nonresonant oscillations with an active neutrino species $\nu_a$ though mass mixing \cite{Dodelson:1993je}, with a mixing angle as small as $\theta\sim 10^{-6}$  for $m_{\nu_s}\sim 100\,$keV \cite{Abazajian:2017tcc}.  
However, these values of active-sterile mixing are robustly excluded by astrophysical observations.
In particular, X-ray searches for the decay $\nu_s\to\nu_a\gamma$, which arises at one loop due to the mixing,
 rule out $m_{\nu_s}
\gtrsim 3\,$keV \cite{Boyarsky:2018tvu}. This forces nonresonantly produced $\nu_s$ into the warm DM regime that is strongly excluded by satellite counts~\cite{Horiuchi:2013noa,Lovell:2015psz,Schneider:2016uqi,Cherry:2017dwu} and Lyman-$\alpha$
constraints~\cite{Boyarsky:2008xj,Viel:2013fqw,Garzilli:2015iwa,Baur:2015jsy,Yeche:2017upn,Garzilli:2018jqh}.

These constraints are relaxed in the presence of a large primordial lepton asymmetry, which can 
allow for resonant oscillations
\cite{Shi:1998km,Dolgov:2002ab,Serpico:2005bc,Laine:2008pg,Boyarsky:2009ix,Canetti:2012kh,Venumadhav:2015pla} in the early universe.
But even in this case the allowed window of parameter space is small \cite{Perez:2016tcq}, considering
the constraints from Big Bang Nucleosynthesis (BBN)
on the lepton asymmetry and from structure formation~\cite{Abazajian:2001nj,Bezrukov:2006cy}.
Various alternative ways of producing the
required $\nu_s$ abundance have been explored,
including decays of Higgs singlets \cite{Kusenko:2006rh,Petraki:2007gq,Adhikari:2016bei}, scenarios in which the $\nu_s$ population is colder relative to production through generic
oscillations \cite{Kusenko:2010ik}, freeze-in~\cite{Datta:2021elq,Das:2021nqj}, through decays of a MeV-GeV scale vector boson \cite{Shuve:2014doa}, or through oscillations modified by neutrino self-interactions~\cite{DeGouvea:2019wpf,Kelly:2020pcy} or scalar fields~\cite{Bezrukov:2018wvd,Bezrukov:2019mak}.

Here we propose a new mechanism for achieving the desired $\nu_s$ abundance through resonant oscillations, by coupling the active neutrinos to a very light dark photon, with a mass as low as $m_{A'}\sim 10^{-12}\,$eV. 
The presence of a primordial condensate of such vector boson particles is shown to modify the dispersion relation of the neutrinos in the early universe.
The leading effect is an enhancement of the active neutrino self-energy, which enables a level crossing with its heavier sterile counterpart and thus facilitates resonant oscillations between the two states.
The dilution of the vector boson condensate caused by the expansion of the universe eventually shuts off the resonance at late times.

The formation of bosonic condensates\footnote{Here, the word condensate should be understood as a coherent classical field configuration and not necessarily as a Bose-Einstein condensate in the quantum mechanical sense.} is conjectured to be possible in the early universe.
The most well-known examples are those of the QCD axion~\cite{Preskill:1982cy,Abbott:1982af,Dine:1982ah} and ultralight scalar dark matter~\cite{Hui:2016ltb}, but an analogous phenomenon is possible for light vector bosons~\cite{Nelson:2011sf}.
The condensate can be described in terms of the classical bosonic field, which acquires a large primordial vacuum expectation value (VEV) that is initially frozen by Hubble friction.
Once the Hubble scale $H$ becomes smaller than the field's mass, the field starts oscillating around the minimum of its potential, the condensate thereafter diluting like a pressureless, non-relativistic fluid.

There are multiple ways in which a vector condensate can be produced in a cosmological environment.
The misalignment mechanism, which is most well-known in its scalar version~\cite{Preskill:1982cy,Abbott:1982af,Dine:1982ah}, fails to generate a VEV in the vector case unless direct couplings to the curvature are present~\cite{Arias:2012az,AlonsoAlvarez:2019cgw}.
Any vector field is however unavoidably sourced through quantum fluctuations during inflation~\cite{Graham:2015rva,Ema:2019yrd,AlonsoAlvarez:2019cgw,Ahmed:2020fhc,Kolb:2020fwh}, producing a cosmological population that can at late times be effectively described by a nonzero VEV.
Finally, an effectively classical vector field configuration can originate from the decay of a dark sector (pseudo)scalar~\cite{Agrawal:2018vin,Co:2018lka,Co:2018lka,Bastero-Gil:2018uel,Long:2019lwl,Bastero-Gil:2021wsf}.

An important question regards the polarization of the dark photon field.
While the misalignment mechanism singles out a preferred direction for the vector field in the whole observable universe, other production mechanisms may produce a field that is originally only locally porlarized, or even completely unpolarized.
Furthermore, the cosmological evolution of such polarization has to date not been studied.
In view of these uncertainties regarding the late-time structure of the dark photon field, in this work we opt to use a simple single-direction description as is commonly done in dark photon studies (see~\cite{Caputo:2021eaa} for a study of the relevance of polarization in phenomenological applications).

The remainder of this paper is structured as follows. Firstly, we describe how nonstandard interactions of active neutrinos with a background vector condensate can resonantly produce $\nu_s$ in Section \ref{sectII}.  In Section \ref{sectIII}, the identification of this vector with a $U(1)_{L_\mu-L_\tau}$ gauge boson is explained, along with the numerous laboratory and cosmological constraints on its couplings. Detailed predictions for the $\nu_s$ relic density are made in Section \ref{sectIV}.  Then, a preliminary study is made in Section \ref{sectV} of the impact of the dark photon dynamics at the present time on active neutrino oscillations.  A summary and conclusions are given in Section
\ref{sectVI}.

\section{Resonant oscillation mechanism}
\label{sectII}
A necessary condition for resonant $\nu_a-\nu_s$ oscillations to occur is for the the matter potential of the active neutrinos $\nu_a$ to be increased and cross the sterile neutrino mass threshold (here and in what follows we always assume that $m_a \ll m_s$).
For simplicity, here we illustrate the situation for sterile mixing with a single active flavor. In practice and due to the smallness of the involved mixing angles, the more general results for mixing with multiple flavors can be easily deduced by adding the individual single-flavor contributions. The Schr\"odinger equation describing the oscillations is governed by the
Hamiltonian
\be 
    H_{as} = p+ \frac{1}{2p}\left({m_{aa}^2\atop m_{as}^2}\, {m_{as}^2\atop m_{ss}^2}\right)
    +\left({V_a\atop 0}\,{0\atop 0}\right)
\ee
in the $(\nu_a,\nu_s)$ flavor basis and in the relativistic limit. Here, $p$ is the neutrino momentum and $V_a$ is the corresponding matter potential.  Resonant flavor transitions occur when $2p V_a + m_{aa}^2 = m_{ss}^2 \cong m_s^2$, indicating the need for a large positive contribution to $V_a$ which does not exist within the standard model (SM). In fact, the electroweak (EW) matter potential in the absence of lepton asymmetries, which we denote by $V_\W$, is negative and given by
\be
    V_\W \cong
    - \frac{14\pi}{45\alpha}(3-s_\W^2)s_\W^2\,G_\F^2 T^4 p  \equiv -C_{\W_{\phantom{|}}\!\!}^2 T^4 p,
    \label{eq:vweq}
\ee
at temperatures $T$ below the electroweak phase transition temperature $T_{EW}$ but higher than the mass of the corresponding charged lepton~\cite{Notzold:1987ik}. Here, $s_\W = \sin\theta_\W$ is the sine of the Weinberg angle and we have
taken the neutrino to be approximately on shell ($E\cong p$) to remove the energy-dependence from
the general expression for the self-energy.
To accurately treat the system at all temperatures and momenta, rather than using the limiting expression for $T\ll T_{EW}$ in (\ref{eq:vweq}), we take the general form of $V_\W$ described in appendix \ref{nuse} in the following numerical treatment.  

\begin{figure}[t]
\begin{center}
\centerline{\!\!\!\!\!\!\!\includegraphics[scale=0.49]{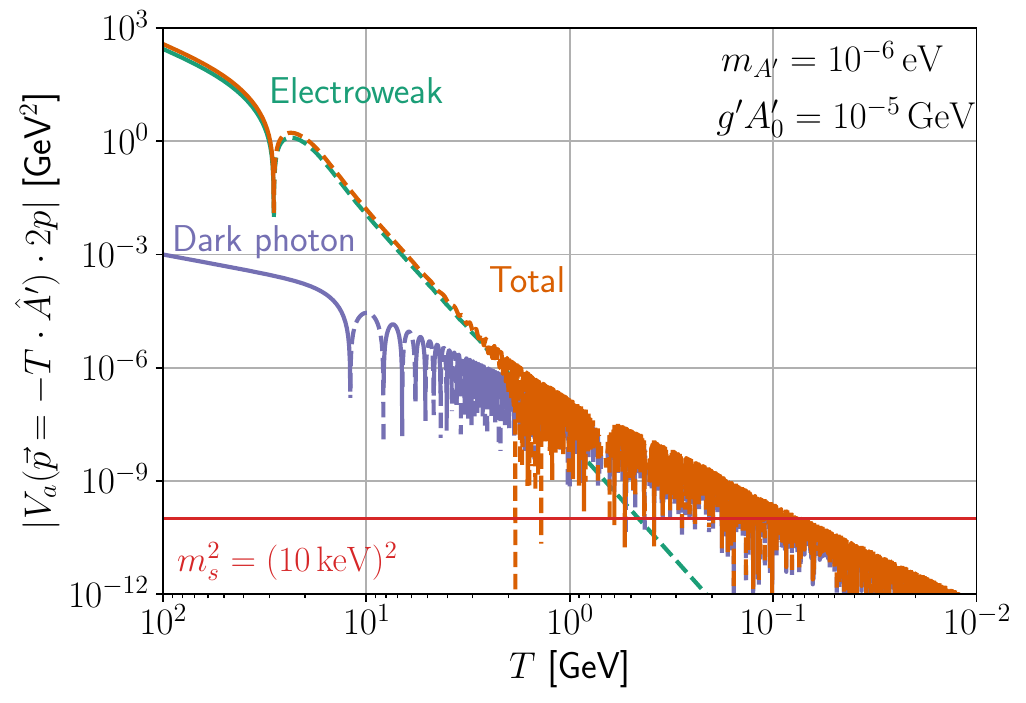}}
 \caption{Absolute value of $V_a$,  the matter potential for active neutrinos, times $2p$, twice the neutrino proper momentum, as a function of the temperature of the universe.
 The green line shows the EW potential in Eq.~\eqref{eq:vweq}, the purple line corresponds to the one induced by the dark photon condensate, Eq.~\eqref{eq:dVa}, and the orange line shows the sum of both contributions.
 Solid (dashed) lines indicate where $V_a$ or its components are positive (negative).
 Resonances occur when $2p V_a$ crosses $m_s^2$, the square of the sterile neutrino mass (drawn as a red line for an exemplary value of $m_s=10$~keV).
 For this figure, parameters were fixed at  $m_{A^\prime}=10^{-6}\,\mathrm{eV}$ and $g' A_0' =10^{-5}\,\mathrm{GeV}$, with the neutrino momentum taken to be antiparallel to the vector VEV, $\vec{\,p} = -T\,\hat{A'}$.  $A'_0$ denotes the initial amplitude of the dark photon field before it starts oscillating at $H\sim m$.}
 \label{fig:Effective_potential}
\end{center} 
\end{figure}

A positive contribution to $V_a$ can arise
by coupling the active neutrino to a new bosonic field with a nonvanishing VEV.  SU(2)$_L$ gauge invariance motivates us to consider a vector boson
$X'_\mu$ with the gauge coupling
\be
{\cal L} = \bar L_a \left(i\slashed{\partial} - g'\slashed{X}'
-m_a\right)L_a\,,
\ee
where $L_a$ is the lepton doublet.  Assuming that the VEV is spacelike and using the gauge freedom to set $\langle X^0\rangle=0$, the dispersion relation in the absence of mixing and in the relativistic limit is
\be
    E \cong p + V_\W +\frac{1}{2p}\left((g'A')^2 -
    2 g' \vec p\cdot \vec A' + m_a^2\right),
\ee
where $A' = X'/a$ is magnitude of the physical vector field, $a$ is the cosmological scale factor, and $p$ is the neutrino proper momentum \cite{Arias:2012az}.
From this expression we can directly read off the new contribution to the matter potential, which becomes $V_a= V_\W + \delta V_a$, with
\be
    \delta V_a = {1\over 2p}\left((g'A')^2 -
    2 g' \vec p\cdot \vec A'\right) \,.
    \label{eq:dVa}
\ee
The total matter potential for an exemplary choice of model parameters is shown in Fig.~\ref{fig:Effective_potential}.
The time evolution of the dark photon field is obtained from its classical equation of motion
\be
    \ddot{A'} + 3 H(t) \dot{A'} + m_{A'}^2 A' = 0,
\ee
which is solved by
\be
    A'(t) \cong 1.08\, A_0'\,{ J_{1/4}(m_{A'}t)\over (m_{A'} t)^{1/4}}\,,
\ee
where $J_{1/4}$ is a Bessel function of fractional order, and
we denote by $A'_0$ the amplitude of the field at early times, when $A'$ is still frozen by Hubble damping. 
As can be seen in Fig.~\ref{fig:Effective_potential}, the condition for resonance, $V_a= m_s^2/2p$, can occur when the the dark photon contribution almost exactly cancels the electroweak one, {\it i.e.}, $\delta V_a \cong -V_\W$, or when the vector potential oscillates at temperatures $H(T)\lesssim m_{A'}$.

\section{Coupling the dark photon to neutrinos}
\label{sectIII}
To couple $A'$
to active neutrinos in an anomaly-free way, the simplest phenomenologically viable choice is to gauge $L_\mu-L_\tau$.  Although $L_e-L_\tau$
and $L_e-L_\mu$ are also anomaly free, there are very stringent constraints on fifth forces, such as that mediated by a light
$A'$ coupling to electrons \cite{Wise:2018rnb}, which would rule out couplings $g'$ down to the $10^{-25}$-$10^{-20}$ level.
In the $L_\mu-L_\tau$ case, fifth forces between the muons present in neutron stars can alter the orbital period and the GW emission of neutron star binaries~\cite{KumarPoddar:2019ceq,Dror:2019uea}, leading to strong bounds, which however only apply to very light ($m_{A'}\lesssim 10^{-10}\,\mathrm{eV}$) mediators.
We therefore focus on $L_\mu-L_\tau$ gauge bosons with masses $m_{A'}\lesssim 1\,\mathrm{eV}$ in what follows.

The U(1)$_{L_\mu-L_\tau}$ gauge symmetry forbids mixing between the active neutrinos in the absence of spontaneous symmetry breaking.
It is therefore necessary to introduce extra scalars charged under the symmetry that can generate the observed neutrino mixing pattern after acquiring suitable VEVs.
Many proposals in this direction have been put forward involving extra Higgs doublets~\cite{Ma:2001md}, soft-breaking terms~\cite{Choubey:2004hn} and/or right-handed neutrinos that are also assumed to carry $L_\mu-L_\tau$ charge \cite{Heeck:2011wj,Asai:2017ryy,Araki:2019rmw}.
The common element of all these models is the presence of U(1)$_{L_\mu-L_\tau}$ symmetry-breaking scalars, which inevitably contribute to the $L_\mu-L_\tau$ dark photon mass.

\subsection{Constraints on the $L_\mu-L_\tau$ gauge coupling}

\begin{figure}[t]
\begin{center}
 \includegraphics[width=\linewidth]{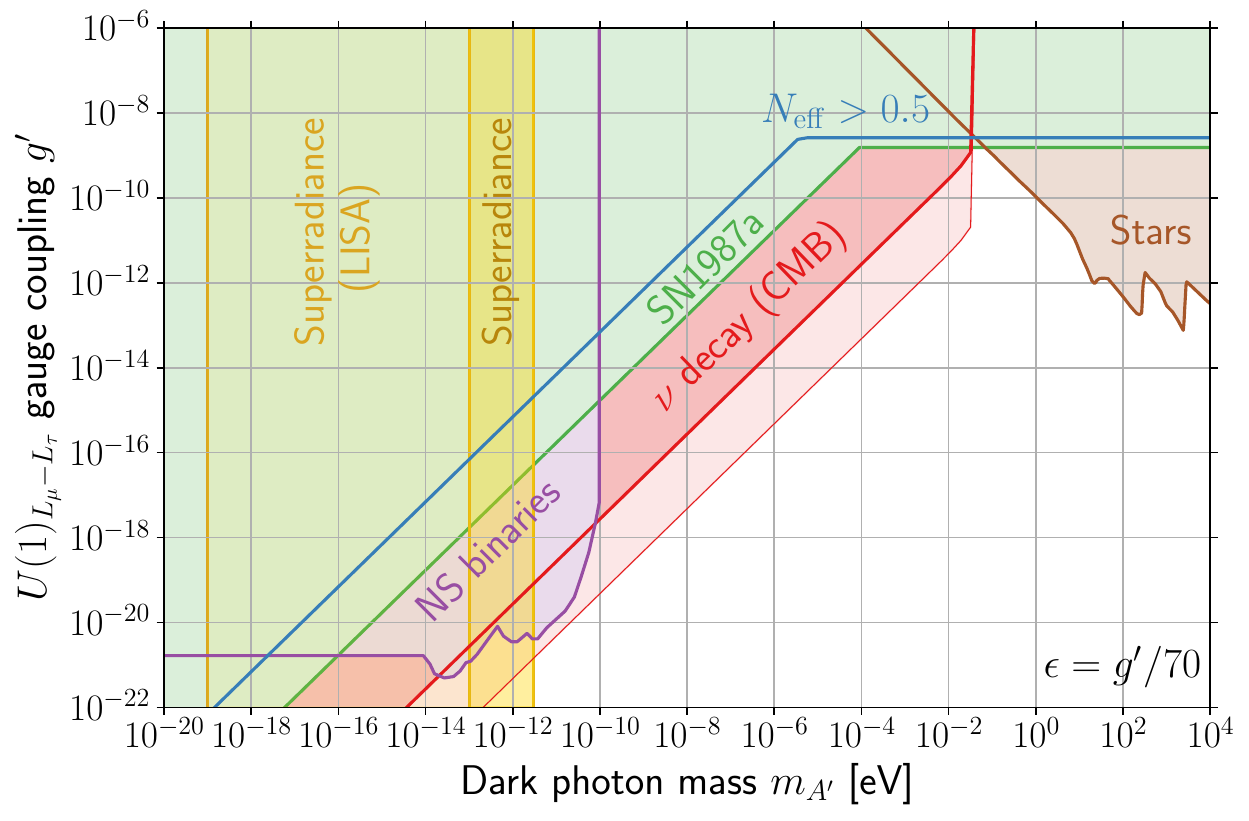}
 \caption{Constraints on a $U(1)_{L_\mu-L_\tau}$ gauge boson in the mass vs. gauge coupling plane, including the loop-induced kinetic mixing with $\epsilon=g^\prime/70$~\cite{Kamada:2015era}.
 Bounds are shown from fifth forces on neutron star binaries~\cite{Dror:2019uea} (purple), modifications of $N_{\rm eff}$ during BBN~\cite{Huang:2017egl,Escudero:2019gzq,Dror:2020fbh} (blue), observations of supernova SN1987A~\cite{Farzan:2002wx,Croon:2020lrf} (green), imprints of neutrino decay in the CMB~\cite{Hannestad:2004qu,Hannestad:2005ex,Escudero:2019gfk,Escudero:2020ped,Barenboim:2020vrr} (red, see text for explanation of the different shades), stellar emission~\cite{An:2013yfc,An:2014twa,Hardy:2016kme,An:2020bxd} (brown), and black hole superradiance~\cite{Baryakhtar:2017ngi,Cardoso:2017kgn,Cardoso:2018tly} (yellow, the lighter band labelled ``LISA'' is a forecast and not a current exclusion). 
 }
 \label{fig:parameter_space_dark_photons}
\end{center} 
\end{figure}

A variety of astrophysical constraints on $g'$ are shown in Fig.~\ref{fig:parameter_space_dark_photons}.
First, the duration of the observed neutrino burst from supernova 1987A constrains the amount of energy that could have been carried away by $A'$ from the supernova.
The abundance of $\mu^\pm$ and $\nu_\mu$ in the proto-neutron star allows a sufficiently strongly coupled ${L_\mu-L_\tau}$ gauge boson to be copiously produced and deplete energy from the exploding star.
The  production  of $A'$ is dominated by semi-Compton scattering for $m_{A'}\sim 10^{-4}\,\mathrm{eV}$~\cite{Croon:2020lrf} and by neutrino pair-coalescence for lighter masses~\cite{Farzan:2002wx}.
The constraint strengthens with decreasing $m_{A'}$ (green region of Fig.~\ref{fig:parameter_space_dark_photons}) because the coupling between neutrinos and the longitudinal dark photon mode scales as $g'm_\nu / m_{A'}$ at energies $E \gg m_{A'}$.
This is a consequence of the gauge current nonconservation in the presence of spontaneous symmetry breaking, as is necessary to allow for $\nu_\mu-\nu_\tau$ oscillations, and can be understood using the Goldstone boson equivalence theorem~\cite{Dror:2020fbh}.

Similar conditions occur in the early universe, where neutrino interactions with the dark photon can alter the effective number of degrees of freedom ($N_{\rm eff}$) and spoil the light-element abundance predictions of Big Bang Nucleosynthesis.
For masses $m_{A'}\gtrsim 4\times 10^{-6}\,\mathrm{eV}$
and couplings $g'\gtrsim 4\times 10^{-9}$, thermalization through $\mu^+\mu^-\leftrightarrow \gamma A'$  leads to an increase in $N_{\rm eff}$~\cite{Escudero:2019gzq}.\footnote{This is only possible if the universe reheats above the muon threshold ($T_{\rm R}> m_{\mu}$); otherwise $A'$ production relies on the interaction with neutrinos~\cite{Huang:2017egl}, which yields bounds that are about $3$ orders of magnitude weaker for high-mass dark photons.}
At smaller $m_{A'}$, the enhanced interaction of the longitudinal component of the vector with neutrinos makes $\nu\nu\to A'A'$ the most efficient process to produce dark photons in the early universe~\cite{Dror:2020fbh}, yielding
stronger limits with decreasing $m_{A'}$, shown by the blue line in Fig.~\ref{fig:parameter_space_dark_photons}.
For sufficiently large $m_{A'}$ and cosmological dark photon abundances, $A'\rightarrow\bar{\nu}\nu$ decays could be an extra source of radiation that contributes to $N_{\rm eff}$\cite{Krnjaic:2020znf}.
Although these early-universe limits are currently weaker than those from SN1987A, they have the potential to significantly 
improve, in light of the expected sensitivity of next-generation Cosmic Microwave Background (CMB) experiments to deviations in $N_{\rm eff}$ at the $0.03$ level~\cite{Abazajian:2016yjj}.

The stronger coupling of the longitudinal $A'$ polarization also enhances the rate of active neutrino decays $\nu_i\rightarrow\nu_j+A'$ for $m_{A'}\ll m_{\nu_i}$,
which can distort the CMB angular and frequency spectra~\cite{Hannestad:2004qu,Hannestad:2005ex}.
A reevaluation of this effect in light of Planck 2018 data~\cite{Escudero:2019gfk} obtained the limit $\tau_{\nu_i} \gtrsim 10^9\,\mathrm{s}\,(m_{\nu_i}/50\,\mathrm{meV})^3$,  strongly constraining $L_\mu-L_\tau$ models~\cite{Escudero:2020ped}.
This was recently disputed in~Ref.\ \cite{Barenboim:2020vrr}, which found a much weaker constraint $\tau_{\nu_i} \gtrsim (4\times10^5-4\times 10^6)\,\mathrm{s}\times(m_{\nu_i}/50\,\mathrm{meV})^5$.
In view of this discrepancy, we conservatively impose the weaker constraint, corresponding to the dark red region in Fig.~\ref{fig:parameter_space_dark_photons}, while the lighter shading depicts the stronger limit.

Beyond the direct limits on $g'$, loop processes involving $\mu$ and $\tau$ leptons induce kinetic mixing between $A'$ and $\gamma$ of strength $\epsilon\sim g^\prime/70$~\cite{Kamada:2015era}.
This allows $A'$ to couple to electrons, 
giving additional constraints from stellar emission of dark photons \cite{An:2013yfc,An:2014twa,Hardy:2016kme,An:2020bxd} that become relevant for larger masses $m_{A'}\gtrsim 0.1\,\mathrm{eV}$.
Finally, observable consequences of the superradiance of spinning black holes triggered by a light vector can rule out the existence of light bosons in a given mass window~\cite{Baryakhtar:2017ngi,Cardoso:2017kgn,Cardoso:2018tly}, independently of their nongravitational interactions.\footnote{Interactions of the dark photon with the plasma around astrophysical black holes may lead to a weakening or nullification of the constraints~\cite{Blas:2020kaa,Cannizzaro:2020uap,Caputo:2021efm}.}\ \
Fig.~\ref{fig:parameter_space_dark_photons} displays the limit from Ref.~\cite{Cardoso:2018tly} in dark yellow, while the lighter yellow region indicates the range of $m_{A'}$ that could be probed by future GW observatories like LISA~\cite{2017arXiv170200786A}. 

 Fig.~\ref{fig:parameter_space_dark_photons} shows that the CMB constraints on decaying neutrinos require $g'$ to be very small for dark photons in the mass region of interest.
However, even such small couplings are easily compatible with an 
initial $g'A_0'$ that can copiously produce sterile neutrinos by the mechanism that we propose.
To characterize the allowed range of $g'A_0'$ values, we note that the energy density of the vector condensate $\langle A'\rangle$ redshifts like dark matter once the field starts oscillating at a temperature $T_{\mathrm{osc}}$.
The relic density in $\langle A'\rangle$ is related to its mass and the initial field value by~\cite{Arias:2012az}
\begin{equation}
    \frac{\Omega_{A'}}{\Omega_{\rm DM}} \simeq 5\,\mathcal{F}(T_{\rm osc}) \left( \frac{A'_0}{10^{12}\,\mathrm{GeV}} \right)^2 \sqrt{\frac{m_{A'}}{\rm eV}},
\end{equation}
where $\mathcal{F}(T_{\rm osc}) = [g_\star(T_{\rm osc})/3.36]^{3/4}/[g_{\star S}(T_{\rm osc})/3.91]$ is an $\mathcal{O}(1)$ function encoding the dependence on the number of relativistic degrees of freedom.
For a given value of $m_{A'}$, the condition that the relic density of dark photons does not exceed a given fraction of $\Omega_{\rm DM}$ sets a limit on the size of the initial $A_0'$.
Combining this with the previously described limits on $g'$ constrains
\be \label{eq:gA_mA_limit}
g'A_0' < 0.5\,\mathrm{GeV} \frac{1}{\sqrt{\mathcal{F}(T_{\rm osc})}} \left( \frac{\Omega_{A'}}{\Omega_{\rm DM}} \right)^{1/2} \left( \frac{m_{A'}}{10^{-6}\,\mathrm{eV}} \right)^{3/4},
\ee
which is valid for $10^{-10}\,\mathrm{eV} \lesssim m_{A'}\lesssim 10^{-2}\,\mathrm{eV}$, for which the CMB constraint on $\nu$ decays dominates the limits on the size of $g'$ (\emph{c.f.} Fig.~\ref{fig:parameter_space_dark_photons}).

Several mechanisms have been proposed to seed the initial dark photon abundance.
Perhaps the simplest and most general one is the gravitational production of longitudinal modes during inflation~\cite{Graham:2015rva}, which yields a relic density of dark photons that depends on the Hubble rate during inflation as
\be
    \frac{\Omega_{A'}}{\Omega_{\rm DM}}\simeq \sqrt{\frac{m_{A'}}{6 \times 10^{-6}\,\mathrm{eV}}}\left( \frac{H_I}{10^{14}\,\mathrm{GeV}} \right)^2,
\ee
and therefore relies on high-scale inflation to produce a sizeable population of dark photons.
However, as we discuss next, these high inflationary scales can be in tension with the mechanisms that can generate a mass for the dark photon, thus favoring alternative production mechanisms~\cite{Agrawal:2018vin,Co:2018lka,Co:2018lka,Bastero-Gil:2018uel,AlonsoAlvarez:2019cgw,Long:2019lwl,Bastero-Gil:2021wsf}.

\subsection{Higgs and St\"uckelberg masses}
\label{sect:stuck}
For U(1) theories, it is consistent to explain 
a gauge boson mass using either the 
Higgs or the St\"uckelberg mechanism. 
The generation of the observed neutrino mixing matrix implies that at least part of $m_{A'}^2$ should arise from the Higgs mechanism, but it is logically possible that both mechanisms contribute, and the latter could dominate.
Although the low-energy phenomenology is equivalent for either case, important differences arise when the UV completions of the model are considered.

For the case of a Higgsed vector, the manifestly gauge-invariant Lagrangian can be written as
\be
    \mathcal{L_{\rm Higgs}} = \left| \left(\partial_\mu + ig'A'_\mu\right) \phi \right|^2 + \lambda \left( |\phi|^2 - \frac{m_\phi^2}{2\lambda} \right)^2,
\ee
where $\phi$ is the Higgs field with mass $m_\phi$ and quartic self-coupling $\lambda$.
At temperatures below $T_{\rm SSB}\sim m_\phi / \sqrt{\lambda}$, spontaneous symmetry breaking (SSB) occurs and the scalar field acquires a VEV, $v_\phi = m_\phi / \sqrt{2\lambda}$.
This in turn generates a mass for the gauge boson, $m_{A'} = g'v_\phi$.
Note that $v_\phi$ is completely determined by $g'$ and $m_{A'}$, but freedom still remains to vary the Higgs mass and self-coupling as long as the ratio $m_\phi / \sqrt{2\lambda}$ is kept fixed.

As an example, for the benchmark values $m_{A'} = 10^{-6}$~eV and $g'\gtrsim10^{-14}$, motivated by the discussion in the next section, one finds $T_{\rm SSB} \sim v_\phi = m_{A'}/g' \gtrsim 100$~MeV.
As we will see, the resonant neutrino oscillations occur at temperatures $T\ll 1\,\mathrm{GeV}$ so that this late epoch of SSB is not an issue.
Nevertheless, such low values of $T_{\rm SSB}$ mean that the dark photon is  massless during high-scale inflation, which is characterized by the Gibbons-Hawking temperature $T_{\rm GH}\sim H_I/2\pi$.
This is the case for most of the relevant parameter space shown in Fig.~\ref{fig:parameter_space_dark_photons}.
As a consequence, the production of dark photons via inflationary fluctuations is largely incompatible with the Higgs mechanism as the origin of the dark photon mass.

The other possibility is the St\"uckelberg mechanism, which can be implemented via the Lagrangian
\be
    \mathcal{L_{\rm St\ddot{u}ckelberg}} = \half f_\theta^2 \left(\partial_\mu \theta + g'A'_\mu\right)^2.
\ee
Gauge invariance is recovered by demanding that the St\"uckelberg field $\theta$ shifts appropriately under a gauge transformation.
In this case, the dark photon mass $m_{A'}=g'f_\theta$ is proportional to the decay constant $f_\theta$.
The main difference with the Higgs case is that in the St\"uckelberg model the $m_{A'}\rightarrow 0$ limit lies at an infinite distance in field space, in a supersymmetric context.
It was argued in~\cite{Reece:2018zvv} that this leads to restrictions in the dark photon parameters based on quantum gravity conjectures.
In particular, the swampland distance conjecture~\cite{Ooguri:2006in} postulates that traversing large distances in field space necessarily causes a tower of states to become light and invalidates the effective theory under consideration.
In our case, this is conjectured to occur when approaching the $m_{A'}\rightarrow 0$ limit, motivating the introduction of a UV cutoff of the effective theory,
\be
    \Lambda_{\rm UV} \lesssim \sqrt{\frac{m_{A'}M_{\rm P}}{g'}}.
\ee
To maintain predictivity for the inflationary production of dark photons, we should demand this cutoff to lie above the energy scale of inflation $E_I\sim \sqrt{H_I M_{\rm P}}$.
This places a constraint on the dark photon mass and gauge coupling, $m_{A'}/g' \gtrsim H_I$, which is numerically similar to the one found by demanding that SSB occurs before inflation in the Higgs case.
It is important to note that this argument has a number of possible loopholes: the swampland distance conjecture has not been rigorously proven and it may be possible to model-build it away~\cite{Craig:2018yld}, for example using the clockwork mechanism \cite{Choi:2015fiu,Kaplan:2015fuy}.

\section{Sterile neutrino dark matter}
\label{sectIV}
To compute the abundance of resonantly produced sterile neutrinos, including matter effects, a rigorous approach is to use the density matrix form of the Boltzmann equations \cite{Enqvist:1990ad,Raffelt:1992uj,Sigl:1992fn}.  However, this is computationally demanding and accurate results can be found though a simpler, intuitive procedure
\cite{Kainulainen:1990ds,Babu:1991at,Cline:1991zb}, based on the production rate of $\nu_s$ through
$\nu_a$-$\nu_s$ oscillations.
This rate can be written as
\be
\Gamma_s =  \Gamma_a \cdot \langle P_{a\to s}\rangle
\label{Gseq}
\ee
and is proportional to the rate of elastic $\nu_a$ scattering  below the weak scale \cite{Notzold:1987ik},
\be
\Gamma_a\cong
{7\pi\over 24}G_\F^2T^4 p \equiv C_a^2 T^4 p,
\label{rateeqs}
\ee
and to the time-dependent oscillation probability $P_{a\to s} = \sin^2(\omega t/2)\sin^2(2\theta_m)$. Here, the matter mixing angle is
\be
\sin^2(2\theta_m) \cong {4 \theta_0^2\over
4 \theta_0^2 + \left( 2 p V_a/m_s^2 - 1\right)^2}\,,
\label{mixang}
\ee
where $\theta_0 \cong m^2_{as}/m_s^2$ is the vacuum mixing angle and $\omega  = V_a + (m_a^2-m_s^2)/2p$ is the oscillation frequency.  
To take into account the decoherence effects due to elastic scatterings, we perform the time average
\bea
    \langle P_{a\to s}\rangle &=& \Gamma_a\int \mathop{\diff t}\,e^{-\Gamma_a t}\,P_{a\to s} \nn\\
    &\cong&\frac{1}{2}\,{ \theta_0^2 m_s^4  \over (p V_a-m_s^2/2)^2 + (p\Gamma_a/2)^2 + \theta_0^2 m_s^4}\,.
    \label{Pa2s}
\eea
Note that elastic scatterings suppresses the oscillation probability if their rate $\Gamma_a$ is larger than the oscillation frequency $\omega$.
It is therefore of crucial importance to correctly estimate the rate of active neutrino scattering in the early universe plasma.
While the approximation (\ref{rateeqs}) for $\Gamma_a$ is convenient for analytic estimates, for our numerical analysis we use the more accurate results that have been tabulated in Refs.~\cite{Asaka:2006nq,Ghiglieri:2016xye}.
We find that using the more precise $\Gamma_a$ evaluation can correct the sterile neutrino yield by up to an order of magnitude.

The relative abundance of sterile versus active neutrinos of momentum $p$ at a given temperature is commonly denoted by $R(p,T)$. This quantity satisfies the Boltzmann equation $\diff R/\diff t = \Gamma_s(1-R)$, whose solution is\,\footnote{This expression assumes that $T\gtrsim 5\,$MeV so that the active neutrinos are still in equilibrium and total neutrino number is not conserved. I practice, since we are interested in solutions where $R\ll 1$, this caveat is numerically unimportant even at lower temperatures.}
\bea
    R(p,T) = 1 - \exp\left(-\int_T^\infty {\mathop{\diff T'}\, \Gamma_s\over H(T')T'}\right).
\label{Tint}
\eea
The total abundance of sterile neutrinos
relative to active ones is obtained by the phase space integration
\be
Y_{s/a} = {\int \mathop{\diff^{\,3}p}\, R(p,T) f(p/T)\over 
\int \mathop{\diff^{\,3}p}\,  f(p/T)}\,,
\label{phsp}
\ee
where $f(p/T)$ denotes the Fermi-Dirac distribution function.
Once the relative abundance $Y_{s/a}$ is known, the number density of sterile neutrinos can be written as $n_s = Y_{s/a} n_{\nu_a}$, where the active neutrino abundance is $n_{\nu_a} = (3/77)s$, and $s = 2891\,$cm$^{-3}$ is the current entropy density.
The fractional contribution to the cosmic energy density is thus $\Omega_s = m_s n_s/\rho_{\rm crit}$.

\subsection{Evaluation of the sterile neutrino abundance}
In general, it is
numerically challenging to evaluate the abundance (\ref{phsp}), but 
several simplifications help.  First, since we are interested in small abundances, we can linearize the exponential in (\ref{Tint}) and then reverse the
order of the $p$ and $T$ integrals in (\ref{phsp}).
After this, the probability (\ref{Pa2s}) can be integrated exactly over the angle
$\alpha$ 
between $\vec A$  and $\vec p$ to give
\bea
  \int_{-1}^1 \mathop{\diff\cos\alpha} \,\langle P_{a\to s}\rangle  = 
   {2\theta_0^2 \over ab}\,\tan^{-1}\!\!\left(2 a b\over a^2-b^2 + c^2\right)\! ,
\label{angint1}
\eea
where 
\bea
    a &=&  \sqrt{4\theta_0^2 + \left( \frac{p\Gamma_a}{m_s^2} \right)^2},\quad
      b = 2\,{ g'A' p\over m_s^2,},\nn\\
     c &=&  1-  \left(g'A'\over m_s\right)^2 - 2\, {p V_\W\over m_s^2} . 
\eea
With this, $Y_{s/a}$ and the resulting cosmological abundance can be written as
\bea
    Y_{s/a} &=& \frac{2M_P\theta_0^2}{3\cdot1.66\,\zeta(3)} I_s(m_s,\, m_{A'},\, g'A'_0),\\
    \Omega_s h^2 &\cong& 3.6\, \theta_0^2 \left(m_s\over{\rm keV}\right) \,M_P \, I_s\,,
\eea
where the $2d$ integral
\be \label{eq:2d_integral}
    I_s = \int\mathop{\diff T}\mathop{\diff q} \frac{q^2\,f(q)\,\Gamma_a}{\sqrt{g_\star(T)}\,T^3 a b}  \tan^{-1}\!\!\left(2 a b\over a^2-b^2 + c^2\right)
\ee
must in general be computed numerically.
We can however make use of the narrow width approximation to gain some further analytic understanding.  

In terms of the angle between $\vec p$ and $\vec A'$, the resonance occurs near $p\cos\alpha = m_s^2/|g'A'|$, and is narrow if
\be
   { \Delta(p\cos\alpha)\over p\cos\alpha} = \sqrt{
    (p\Gamma_a/m_s^2)^2 + 4\theta_0^2} \ll 1.
\ee
Since we are only interested in $\theta_0\ll 1$, the condition for narrow resonance is simply $p\Gamma_a\ll m_s^2$.
Given that the phase space integral introduced above is dominated
by $p\sim T$, this implies $G_F T^3 \lesssim m_s$, hence
$T \lesssim 1\,$GeV for $m_s \cong 10\,$keV.
This is satisfied for the parameters of interest here, and allows one to approximate the integral Eq.~\eqref{angint1} by
\be
    \int_{-1}^1 \mathop{\diff\cos\alpha} \,\langle P_{a\to s}\rangle \cong
       2\pi{\theta_0^2 \over |ab|}\,\Theta(|b|-|c|)\,.
\label{angint}
\ee
The $q$-integral in Eq.~\eqref{eq:2d_integral} can then be carried out analytically in terms of polylog functions, but the task of solving for the limits of integration (the roots of $|b|=|c|$) becomes too numerically demanding once the vector field unfreezes and starts oscillating. This eventually makes this procedure more costly than the straight two-dimensional numerical integration.

Nevertheless, the expression~\eqref{angint} is helpful to understand the dependence of the sterile neutrino yield on the vacuum mixing angle $\theta_0$.
The width of the resonance is determined by a competition between the vacuum mixing angle and the active elastic scattering rate, as can be seen in Eq.~\eqref{Pa2s}.
This fact singles out two distinct regimes in the calculation.
For resonances occurring at high $T$ the scattering rate is large, so that typically $2\theta_0\ll p\Gamma_a / m_s^2$. In this case, $a\simeq p\Gamma_a/m_s^2$ and Eq.~\eqref{angint} implies that the abundance of sterile neutrinos scales as $Y_{s/a}\propto\theta_0^2$.
On the other hand, the width of resonances occurring at low $T$ is controlled by the vacuum mixing angle, implying that $a\simeq 2\theta_0$ and therefore $Y_{s/a}\propto \theta_0$.
This expectation is confirmed by the numerical results described below.

The dependencies of the sterile neutrino yield on the other parameters---the $\nu_s$ mass $m_s$, the dark photon mass $m_{A'}$, and the product of the gauge coupling and the initial field value $g'A_0'$---are more complex and their understanding requires a full numerical study, to which we now turn.

\subsection{Numerical results}
\begin{figure}[t]
\centering
\includegraphics[width=0.48\textwidth]{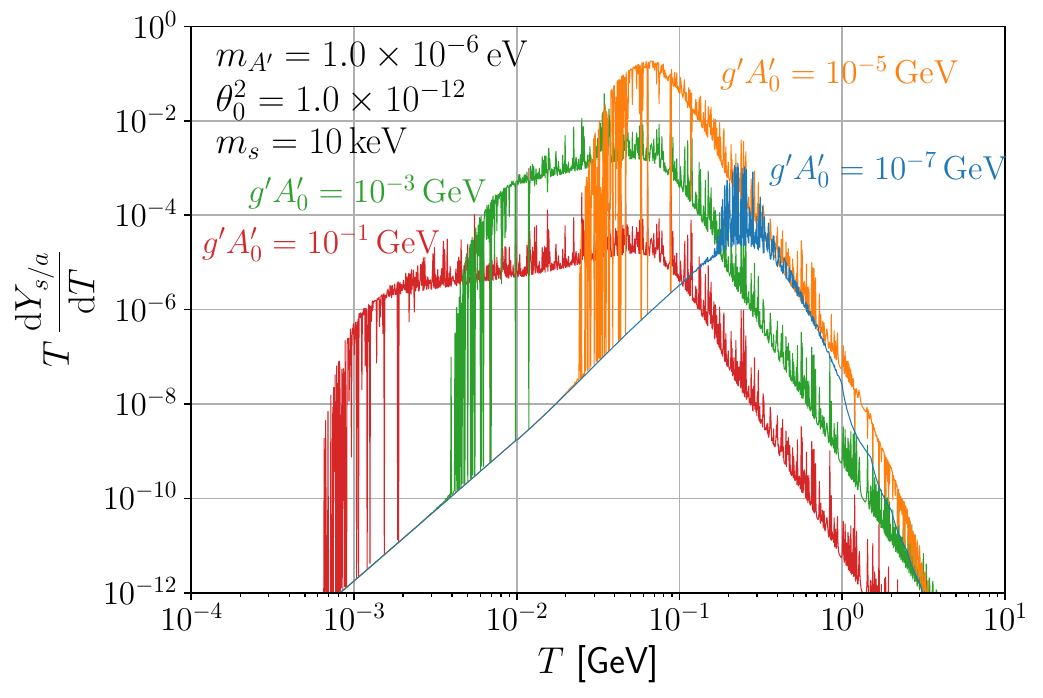}
\caption{Differential sterile neutrino abundance $Y_{s/a}$ produced as a function of temperature,
 for several initial values of $g'A'$, and fixed $m_{A'}$, $\theta_0$, and $m_s$ as indicated in the figure. The vertical jumps in the curves are due to fast oscillations in the dark photon field.}
 \label{fig:dYdT}
\end{figure}

\begin{figure}[t]
\centering
\includegraphics[width=0.47\textwidth]{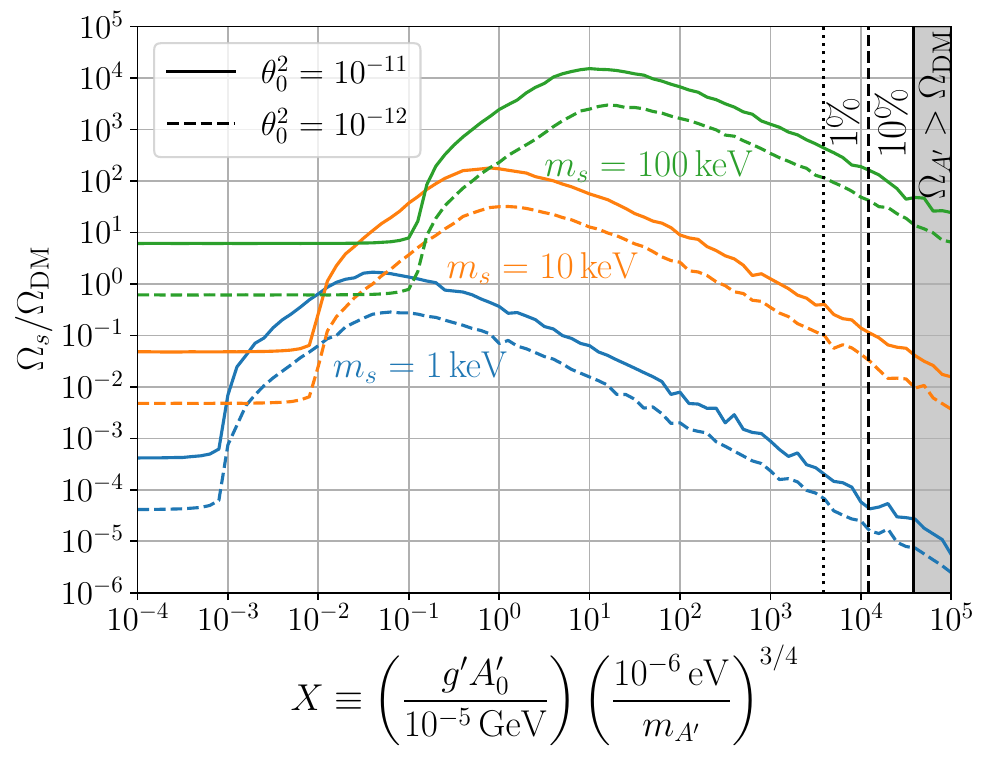}
\caption{Integrated sterile neutrino relic density $\Omega_s$ as a function of the product  $X \propto g'A_0'\cdot m_{A'}^{-3/4}$, for several values of $m_s$ and $\theta_0^2$.
In the area shaded in grey the dark photon overcontributes to the dark matter energy density, the dashed and dotted lines indicate a minimum $10\%$ and $1\%$ contribution given the constraints on $g'$ and $m_{A'}$ shown in Fig.~\ref{fig:parameter_space_dark_photons}.}
 \label{fig:Omega}
\end{figure}

The results of the evaluating Eq.~\eqref{eq:2d_integral} numerically, before and after doing the integration over $T$, are shown in Figs.~\ref{fig:dYdT} and \ref{fig:Omega}.
Fig.~\ref{fig:dYdT} illustrates the temperature evolution of the differential sterile neutrino abundance, demonstrating that bulk of the production occurs at temperatures $T\ll 1\,\mathrm{GeV}$.
At this time the Hubble rate has already fallen below $\sim 0.6\,m_{A'}$, after which the vector field is no longer frozen but is instead oscillating around the minimum of its potential at $A'=0$ \cite{AlonsoAlvarez:2019cgw}.
The temperature at which this transition occurs is
\be
    T_{\rm osc} \equiv \sqrt{0.6\, m_{A'} M_p\over 1.66\,\sqrt{g_*}}\,,
\label{Toscdef}
\ee
and for $T<T_{\rm osc}$, the amplitude redshifts as
\be
    A' \simeq A'_0\left(T\over T_{\rm osc}\right)^{3/2}\,.
\ee
The benchmark value $m_{A^\prime} = 10^{-6}\,\mathrm{eV}$ of the dark photon mass chosen in Fig.\ \ref{fig:dYdT}  corresponds to $T_{\rm osc} \simeq 22$\,GeV, at which point
$g_*= 87.2$ \cite{Laine:2015kra}.
The sterile neutrino production thereafter occurs at the subsequent narrow resonances when the dark photon field value passes through zero and the resonant condition $|b|\geq |c|$ is satisfied.
These events are responsible for the bulk of the sterile neutrino production and are visible as sharp vertical features in Fig.\ \ref{fig:dYdT}.

The basic properties of Figs.~\ref{fig:dYdT} and~\ref{fig:Omega} can be understood in terms of the conversion probability given in Eq.\ (\ref{Pa2s}).  Production of $\nu_s$ is most efficient when the denominator is minimized, which occurs when the resonance condition
$p V_a = m_s^2/2$ is satisfied, and when the full width of the resonance $((p\Gamma_a/2)^2 + \theta_0^2 m_s^4)^{1/2}$
is small.  Since the average momentum is $p\sim 3T$, the latter condition is fulfilled for some optimal temperature
$T_*$ given by 
\bea
    \sfrac32 T_*\, \Gamma_a(T_*) \sim \theta_0\, m_s^2 \;
    \implies \; T_* \sim \left(\theta_0 m_s^2\over C_a^2\right)^{\sfrac{1}{6}}\,.
    \label{Tmaxeq}
\eea
For the fiducial parameters shown in Fig.~\ref{fig:dYdT}, this temperature is $T_*\simeq 0.06\,\mathrm{GeV}$.
It is important to note that the estimate (\ref{Tmaxeq}) is only reliable if resonant conversions are occurring at $T\sim T_*$.
This is only the case if \emph{(i)} the dark photon field has already begun oscillating at that temperature (\emph{i.e.}\! if $T_{\rm osc} > T_*$) and if \emph{(ii)} the amplitude of the oscillations around $T_*$ is still large enough such that the sterile neutrino mass threshold can be overcome, which requires that $g'{A}'(T_*) \gtrsim m_s^2/T_*$.
When ${A}'$ becomes too small to satisfy the resonance condition, sterile neutrino production falls quite abruptly, as Fig.\ \ref{fig:dYdT} shows in the low-$T$ regions.
For small values of $g'A_0'$, like the one corresponding to the blue curve in Fig.~\ref{fig:dYdT}, resonant conversions halt too early to efficiently produce a large sterile neutrino abundance.

Since $\nu_s$ production is determined by ${A}'$ at the time of resonance, the dependence on  dark photon parameters is only through the combination
\be
    X\equiv \left(g'A_0' \over 10^{-5}\,{\rm GeV}\right)  \left(10^{-6}\,{\rm eV}\over m_{A'}\right)^{3/4} \propto \left. g'{A}'\right|_{T>T_{\rm osc}} \,.
\ee 
Hence, in Fig.~\ref{fig:Omega} we choose to show the integrated $\nu_s$ abundance as a function of this quantity.
There, one can see that $\Omega_s$ is maximized at some intermediate value of $X$, which depends on $m_s$
through the resonance condition $g'{A'} \sim m_s^2/T$. One can understand the trends in Fig.\ \ref{fig:Omega} leading to the peaks: at large $X$ the yield is a decreasing function of $X$, since larger 
 $g'A'$ produces a narrower resonance,
 while at small $X$, 
the resonance condition stops being fulfilled before the optimal temperature $T_*$ is reached, resulting in less $\nu_s$ production.

The nonmonotonic nature of the curves in Fig.\ \ref{fig:Omega} implies that there can be two values of
$X$ consistent with $\nu_s$ constituting the observed relic
density.  For sufficiently small $X$, resonant production 
no longer occurs, and instead the nonresonant Dodelson-Widrow mechanism \cite{Dodelson:1993je} dominates.
This corresponds to the flat regions in Fig.\ \ref{fig:Omega}, which are insensitive to the properties of $A'$.  For very large $X$, the $A'$ itself starts to constitute a significant part of the dark matter, leading to the ruled-out grey part of the figure, as well as intermediate regions where $A'$ contributes a subdominant yet significant fraction of the DM.

\begin{figure}[t]
\centering
\includegraphics[width=0.47\textwidth]{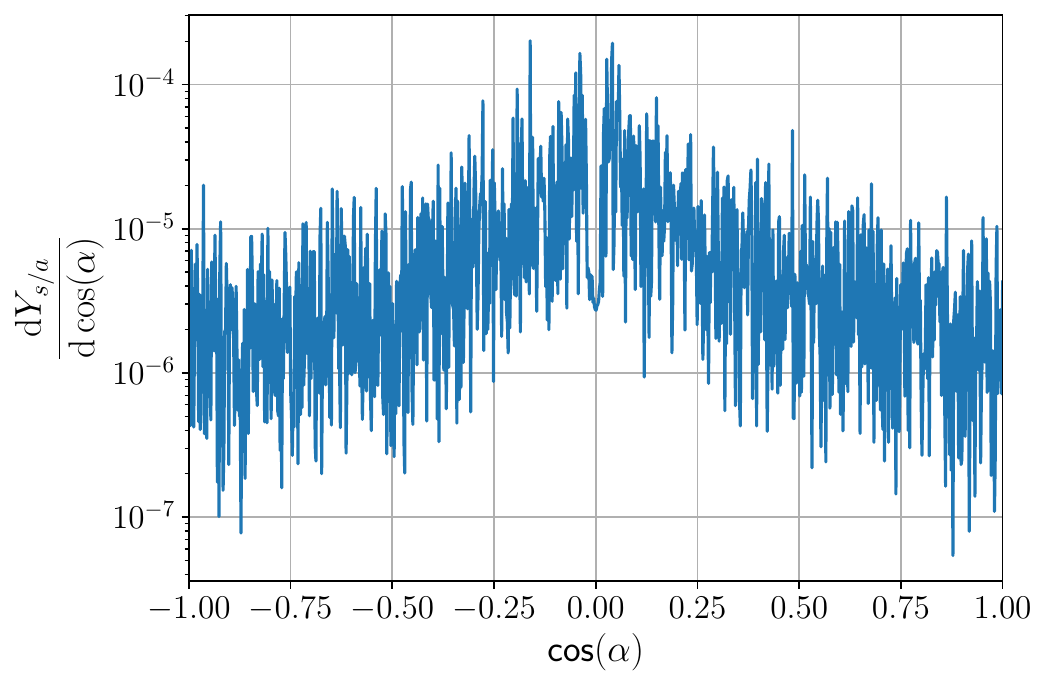}
\caption{Differential sterile neutrino abundance as a function of $\alpha$, the angle between the neutrino momentum $\vec{p}$ and the dark photon vector field $\vec{A}'$. For this figure we use the benchmark parameters $m_s=10$
~keV, $\theta_0^2=10^{-12}$, $m_{A'}=1-^{-6}$~eV, and $g'A'_0=10^{-5}$~GeV. The zigzagging appearance of the curve is a result of the numerical integration over the fast oscillations of the dark photon field.}
 \label{fig:dYsa_dcosalpha}
\end{figure}

It is interesting to study the dependence of the produced sterile neutrino abundance on the angle between the neutrino momentum and the dark photon field.
This is shown in Fig.~\ref{fig:dYsa_dcosalpha}.
The existence of a preferred spatial direction in the form of the vector field VEV translates into an anisotropy in the momentum distribution of the produced sterile neutrinos.
As can be seen in Fig.~\ref{fig:dYsa_dcosalpha}, active-sterile oscillations occur preferentially for momenta almost orthogonal to the direction singled out by 
$\vec{A}'$.
This can be understood qualitatively by reversing the order of integration between $p$ and $\cos\alpha$ in Eq.\ (\ref{phsp})
and using the narrow width approximation to estimate the 
integral over $p$.  The resulting integrand blows up as
$\cos\alpha\to 0$, because the width of the resonance diverges in this limit.  It is the increasing width of the resonance that explains the preference for small $\cos\alpha$.
As $\cos\alpha$ becomes close to zero, the resonance shifts to large neutrino momenta which are Boltzmann suppressed, and we recover the non-resonant (DW) limit for $\cos\alpha=0$.
This behaviour is visible as a sudden dip in the differential abundance shown in Fig.~\ref{fig:dYsa_dcosalpha} around $\cos\alpha=0$.

The fact that the momenta of the DM sterile neutrinos preferentially lie close to the plane perpendicular to $\vec{A}'$ can have  phenomenological consequences.
If, as we are assuming here, the VEV $\vec{A}'$ is uniform throughout the whole universe, the initial velocity dispersion of the DM particles is highly anisotropic.
This could potentially leave an imprint on cosmological observables, in particular for sufficiently light sterile neutrinos which are not completely cold by the start of structure formation.
On the other hand, if $\vec{A}'$ presents variations on sufficiently small scales, this effect averages out and no observable effects are expected.

\begin{figure*}[t]
\begin{center}
\includegraphics[width=0.99\linewidth]{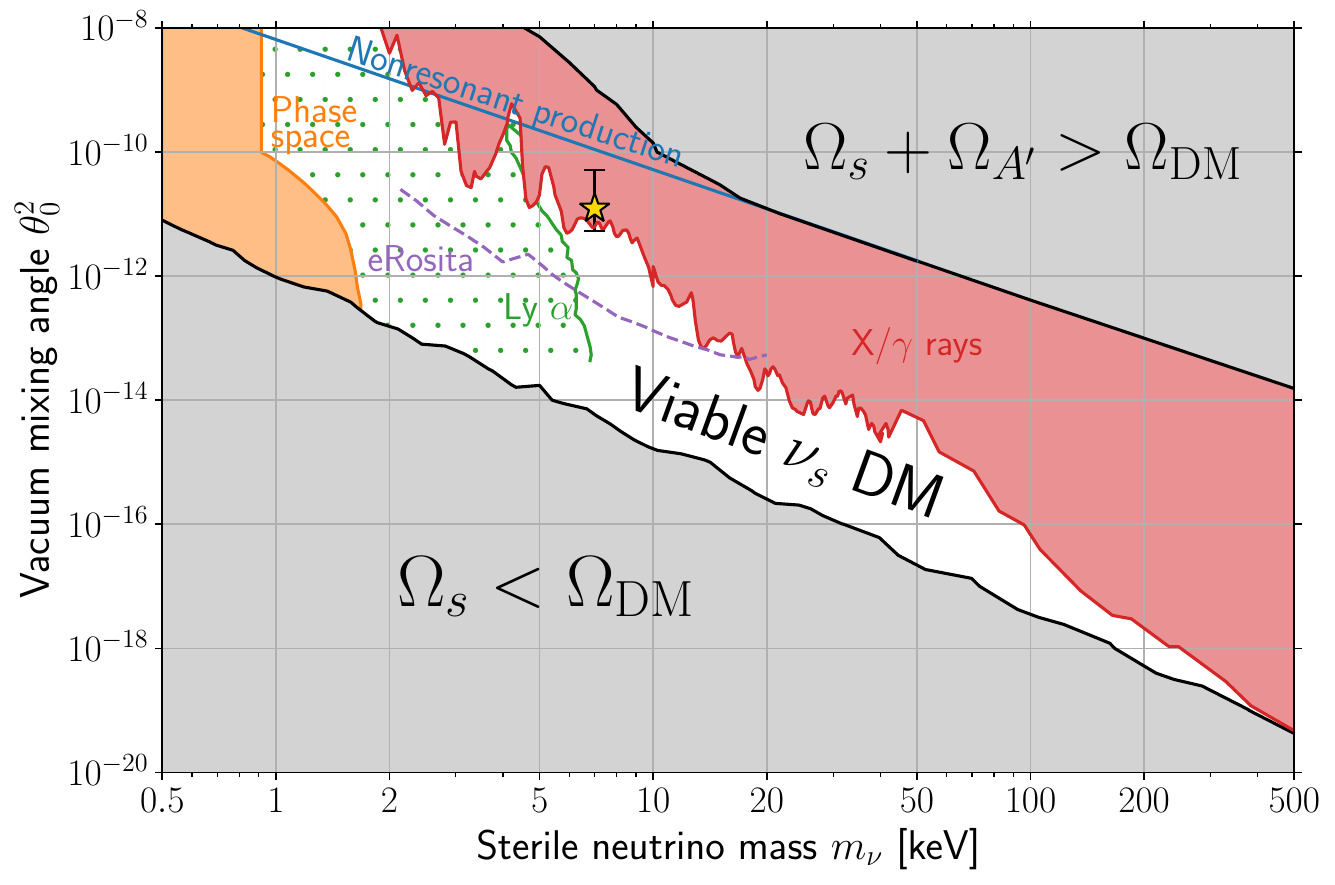}
\caption{Region of the $m_s$-$\theta_0^2$ plane where sterile neutrino dark matter can be produced via dark photon-induced resonant oscillations, corresponding to the unshaded white area.
In the areas shaded in grey, the mechanism either underproduces or overproduces dark matter as explained in the text.
The blue line depicts the line where non-resonant oscillations yield the observed dark matter abundance in the absence of the $L_\mu - L_\tau$ gauge interactions.
We also show constraints coming from $X/\gamma$-ray searches (in red) and prospects for the eROSITA telescope (purple dashed line), as well as the area excluded by phase-space considerations in dwarf galaxies (in orange) and the one disfavored by Lyman-$\alpha$ observations (green dots).
}
\label{fig:cont}
\end{center} 
\end{figure*} 

To project our findings onto the $m_s$-$\theta_0^2$ plane, we marginalize over the dependence of the mechanism on the dark photon parameters, determining at given values of 
$m_s$ and $\theta_0$ the maximum $\nu_s$ abundance that 
can be generated by varying  $g'A_0'$
and $m_{A'}$. 
The result is shown in Fig.~\ref{fig:cont}, in which the 
white  region indicates where our mechanism can successfully generate $\nu_s$ constituting all of the DM.  In the grey areas, the $\nu_s + A'$ combined abundance falls above or below the observed
DM density.  The region excluded by X- and $\gamma$-ray searches~\cite{Essig:2013goa,Horiuchi:2013noa, Tamura:2014mta, Roach:2019ctw} is shown in red, while 
the purple dashed line depicts the potential reach of the eROSITA telescope (we use the prospects from~\cite{Dekker:2021bos}, see also~\cite{Caputo:2019djj,Barinov:2020hiq}).
In the orange region, sterile neutrinos are too light to be bound in the DM halos of dwarf spheroidal galaxies~\cite{Alvey:2020xsk}, and in the green-dotted one they are likely to be inconsistent with observations of the Lyman-$\alpha$ forest~\cite{Baur:2017stq}, although a reevaluation of the constraints using the distorted $\nu_s$ momentum spectrum in our mechanism would be necessary in order to establish a robust limit.
Note that in this region we expect the anisotropy of the sterile neutrino momenta predicted by our scenario to become relevant and potentially observable.

It is worth noting that the contour along which the nonresonant Dodelson-Widrow (DW) mechanism~\cite{Dodelson:1993je} would give the observed relic density, shown in Fig.~\ref{fig:cont}, does not quantitatively agree with previous evaluations in the literature (see, \emph{e.g.}\!\,, Fig.~14 in~\cite{Boyarsky:2018tvu}).
We have recomputed it as a by-product of our analysis, since it corresponds to the $g'\to 0$ limit of our model, finding $\sim 10$ times larger abundance in the high-$m_s$ region relative to previous analyses of DW production~\cite{Boyarsky:2018tvu}. The discrepancy comes from the use of the simplified approximation (\ref{rateeqs}) for the active neutrino damping rate $\Gamma_a$ in previous literature, instead of the more quantitative results from Ref.\ \cite{Asaka:2006nq,Ghiglieri:2016xye} that we have employed.  

Fig.~\ref{fig:cont} clearly showcases that the new mechanism of $A'$-induced resonant production of $\nu_s$ opens the sterile neutrino DM parameter space dramatically.
This is particularly relevant because previously untested regions of it will soon be probed as X-ray searches and structure formation observables continue to grow in sensitivity. Finally, we note that the mechanism can also accommodate a sterile neutrino DM explanation of the $3.5\,\mathrm{keV}$ X-ray line~\cite{Bulbul:2014sua,Boyarsky:2014jta}, marked by the yellow star in Fig.~\ref{fig:cont}.

\section{Active neutrino oscillations}
\label{sectV}
The framework under study can also be tested through its potential impact on laboratory measurements of neutrino oscillations, particularly at long baseline experiments 
sensitive
to solar ($\nu_e\to\nu_\mu$) or atmospheric ($\nu_\mu\to\nu_\tau$) conversions \cite{Fukuda:1998mi}.
This can happen if the new contributions to the neutrino matter potential $\delta V_a$ from the oscillating
$A'$ field are still significant at the present time.
We will present an exhaustive study in the future
\cite{inprep}; here we make a preliminary investigation to show that there is no conflict with the allowed region in
Fig.~\ref{fig:cont}.

To study the effect of the new terms in Eq.\ (\ref{eq:dVa}) in terrestrial neutrino
oscillations, we need the present value $A'_\odot$ of the amplitude of the vector in the solar neighborhood,
which is related to its initial value by
\be
    g'A'_\odot \simeq 1.3\times 10^{-13}{\,\rm eV}\left({10^{-10}{\,\rm eV}\over m_{A'}}\right)^{3/4}\left(g'A_0'\over 1{\,\rm keV}\right)
    \label{gpAprel}
\ee 
due to the redshifting of the energy density of $A'$ before structure formation, and the subsequent concentration of DM in the galaxy, 
 $\rho_{\rm DM}^\odot / \rho_{\rm DM}^{\rm av}\sim 3\times 10^5$.  The value of $g'A'_\odot$ is therefore much smaller than the typical neutrino energy in oscillation experiments,
so that the  $(g'A')^2,$ term in  Eq.\ (\ref{eq:dVa}) can be neglected,
and the Hamiltonian takes the form
\be
   H_{\mu\tau} = p+ {1\over 2p}\left({m_\mu^2\atop m_{\mu\tau}^2}\, {m_{\mu\tau}^2\atop m_\tau^2}\right)
    +g'\vec A_\odot'\cdot \hat p\left({-1\atop \phantom{-}0}\ {0\atop1}\right),
    \label{Hmutau}
\ee
where the vector is oscillating at the present time as $\vec A_\odot'(t) = |\vec A_\odot'|\cos(m_{A'}t)$, and $\hat p$ is a unit vector.
This system can be solved numerically and compared with standard oscillations, leading to constraints on $g'A'_0$ and $m_{A'}$ to avoid too big a deviation.

It is convenient to rescale to a dimensionless time variable $\hat t = 
(\Delta m^2/4p) t$ for each momentum $p$, where 
{$\Delta m^2$ is the atmospheric ($\Delta m_{23}^2 \simeq 2.5\times 10^{-3}$\,eV$^2$) or solar ($\Delta m_{12}^2 \simeq 7.4\times 10^{-5}$\,eV$^2$) mass splitting.}
With this, the Schr\"odinger equation takes the form
\be
    i{d\over d\hat t}\left({\psi_{\nu_\mu}\atop \psi_{\nu_\tau}}\right) \cong \left( {-(\cos2\theta+h)\atop \sin2\theta}\ {\sin2\theta\atop \cos2\theta + h}
        \right)\left({\psi_{\nu_\mu}\atop \psi_{\nu_\tau}}\right),
                \label{atmosc}
\ee
with
\bea
    h(t) &=& 4\, {p|g'A'_\odot|\cos\alpha\over\Delta m^2}\cos\left(
       { 4 p\ m_{A'}\over \Delta m^2}\hat t\right)\nn\\
       &\equiv& h_0 \cos\omega\hat t,
              \label{thatdef}
\eea
where $\alpha$ is the angle between $\vec p$ and $\vec A'_\odot$.
In the $\hat t$ variable, all momentum modes have the same oscillation frequency in the absence of the new physics effects from $h(t)$.

As an example, we consider the case of atmospheric oscillations and approximate the vacuum $\nu_\mu$-$\nu_\tau$ mixing angle as $\theta = 45^\circ$ in Eq.\ (\ref{atmosc}) for simplicity.  
Here we also set $\alpha=0$; a more detailed study will average over the directions of the neutrinos relative to $\vec A'$,
slightly weakening the ensuing constraints.

\begin{figure}[t]
\begin{center}
 \includegraphics[width=0.48\textwidth]{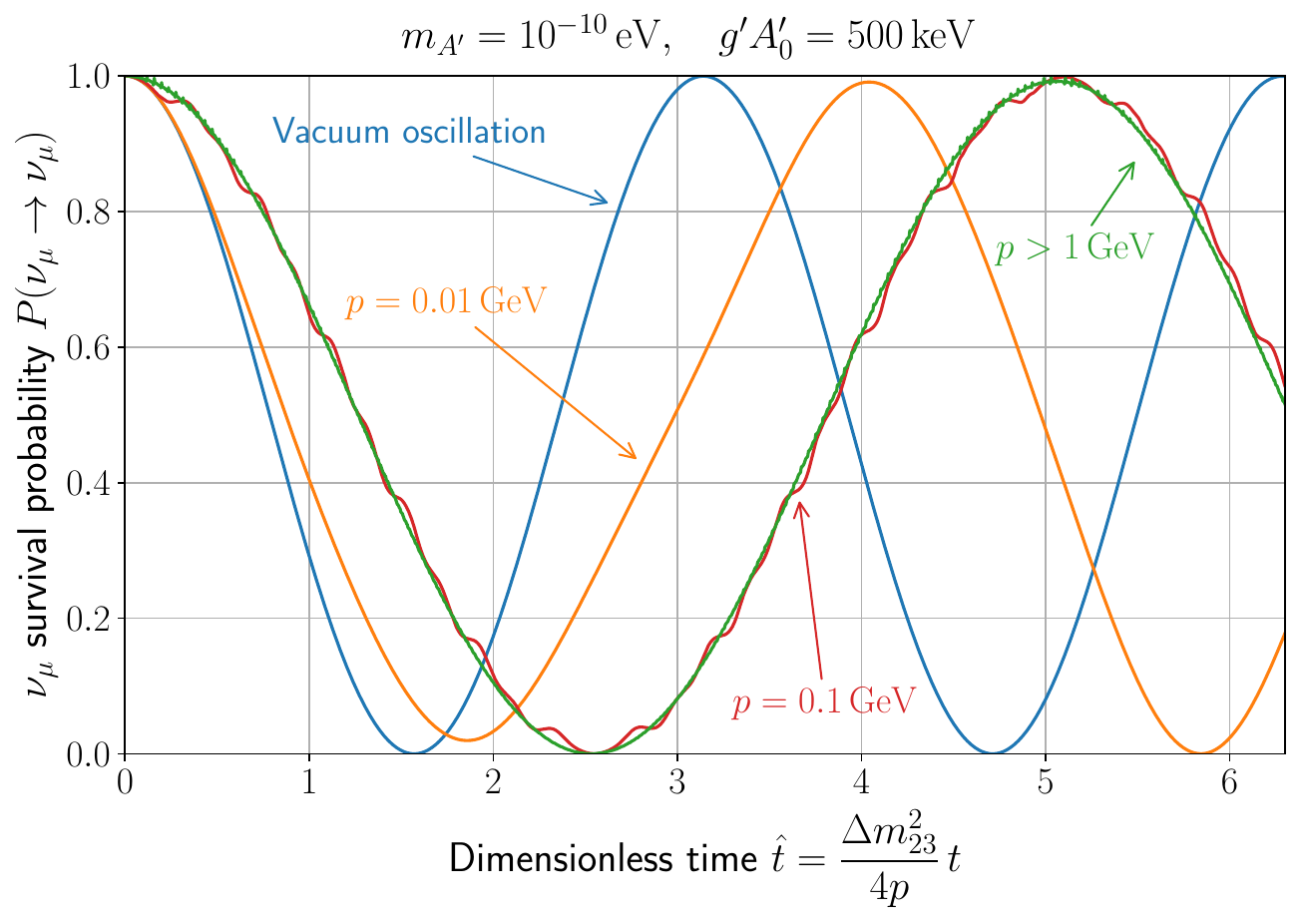}
 \caption{Muon neutrino survival probability in atmospheric neutrino oscillations as a function of the dimensionless time parameter $\hat t$ defined in eq.~\eqref{thatdef}, modified by interactions with an oscillating dark photon background. For this illustrative example, we fix $g'A_0' = 500\,$keV and $m_{A'} = 10^{-10}$\,eV.  Curves are shown for  neutrino momenta $p>1\,$GeV (green), $p=0.1\,$GeV (red), and $p=0.01\,$GeV (orange), as well as the standard prediction  when $g'A' = 0$ (blue) as a reference.
}
 \label{fig:atmnu}
\end{center} 
\end{figure}

Results for a set of parameters for which the new physics effect would be appreciable in
current atmospheric oscillation experiments are shown in Fig.\ \ref{fig:atmnu}.
For small momenta $p\lesssim \Delta m^2 / 4 m_{A'}$, a distortion of the shape of the survival probability curve becomes evident, while for large $p$ these disappear, except for an increase in the oscillation length relative to the standard vacuum oscillation result,
which has the same effect as a decrease in $\Delta m^2$. 

This effect can be understood by using second-order time-dependent perturbation theory to approximately solve Eq.~\eqref{atmosc}. One finds an imaginary secular term growing linearly in time that can be resummed to produce a shift $\delta H$ in the effective Hamiltonian, 
appearing in the evolution operator $\exp(i(H_0 + \delta H)t)$.  The 
effect of $\delta H$ is precisely the same as a multiplicative correction to the vacuum mixing parameter $\Delta m^2$,
\be
    \Delta m^2 \to \Delta m^2 \left(1 - \left(g'A'_\odot \sin 2\theta\over 2 m_{A'}\right)^2\right),
\ee
written here for a general mixing angle $\theta$.
By demanding that the fractional change in
$\Delta m^2$ does not exceed the experimental error~\cite{Esteban:2020cvm} (which may be too stringent since for high $p$ this effect is degenerate with the vacuum neutrino squared mass difference), and using Eq.\ (\ref{gpAprel}), one can derive the limit
\be
   { g'A'_0} \lesssim 170\, {\rm keV}\left(m_{A'}\over 10^{-10}\,{\rm eV}\right)^{7/4}.
   \label{gpAplim}
\ee
Although this bound is strong for small dark photon masses, it degrades steeply with increasing $m_{A'}$ and becomes subleading to the constraint in Eq.~\eqref{eq:gA_mA_limit} for $m_A'\gtrsim 3\times 10^{-10} (\Omega_{A'}/\Omega_{\rm DM})^{1/2}$~eV.
As the production of the $\nu_s$ relic density can always be mediated by dark photons heavier than this threshold, this constraint does not further restrict any of the parameter space shown in Fig.~\ref{fig:cont}.
For lighter dark photons, neutrino oscillations constitute a promising way to indirectly test the sterile neutrino DM paradigm presented in this paper.
Since the experimental error in $\Delta m^2_{23}$ is
relatively smaller than that for $\Delta m^2_{12}$, we find that solar neutrino oscillations give a somewhat less stringent limit than (\ref{gpAplim}).

\section{Summary and Conclusions}
\label{sectVI}
In this work, we have shown how the existence of a cosmological dark photon condensate can catalyze the production of keV sterile neutrino dark matter in the early universe.
If active neutrinos couple to the dark photon, which is assumed to be very light and weakly coupled, their dispersion relation is modified in such a way that resonant conversions into a sterile flavor can take place.
For concreteness, we have implemented the coupling between the dark photon and the neutrinos by gauging the anomaly-free combination $L_\mu - L_\tau$, although other possibilities may be viable.

Our main finding is that sterile neutrinos with masses in the few to $\sim500$ keV range and $\theta_0^2 \sim 10^{-18}-10^{-10}$ mixings, produced through the resonant oscillations enabled by the dark photon, may constitute all the dark matter of the universe. The mass and interaction strength of the dark photon can span a large range, $10^{-12}\,\mathrm{eV}\lesssim m_{A'}\lesssim 1\,\mathrm{eV}$, and $g'\lesssim 10^{-10}$.
    
The new resonant mechanism is much more effective than its nonresonant counterpart, commonly known as Dodelson-Widrow production~\cite{Dodelson:1993je}. This can be seen in Fig.~\ref{fig:Omega}, where the nonresonant yield corresponds to the low plateau of the curves at the left of the figure, and in Fig.~\ref{fig:cont}, where the favored region for $\nu_s$ DM production extends to much lower mixings than the blue line representing the DW prediction.
Compared to another popular mechanism for resonant production of sterile neutrino dark matter \cite{Shi:1998km}, our proposal has the appeal of not requiring the existence of a large primordial lepton asymmetry.
    
The small dark photon mass can in principle be generated by either the Higgs or the St\"uckelberg mechanism. 
We have noted that both these mechanisms are incompatible with the gravitational production of dark photons during high-scale inflation: the St\"uckelberg option is disfavored by the swampland distance conjecture~\cite{Reece:2018zvv}, while the spontaneous symmetry breaking associated with a Higgs mass is only expected to occur much later in the history of the Universe.
As a consequence, dark photon production mechanisms that occur at comparably lower scales, like the ones put forward in~\cite{Agrawal:2018vin,Co:2018lka,Co:2018lka,Bastero-Gil:2018uel,AlonsoAlvarez:2019cgw,Long:2019lwl,Bastero-Gil:2021wsf}, are favored in our proposal.
    
We have also discussed how the existence of a $L_\mu - L_\tau$ dark photon condensate in the universe can impact the oscillation pattern of $\mu$ and $\tau$ neutrinos. This may have interesting implications for long baseline or atmospheric neutrino oscillation experiments. We have only presented a preliminary study of this effect here, and we will analyze it in more detail in an upcoming publication~\cite{inprep}.

Our present study is based on the assumption that the dark photon field presents an overall fixed polarization in the entire universe.
This is the case for some popular production mechanisms, but others make different predictions for the correlation length in the polarization of the field.
Although it would be interesting to extend our analysis to other scenarios, the strong model dependence and the unknown effect of the cosmological evolution on the polarization greatly complicates this task.
Nevertheless, we do not expect the sterile neutrino abundances calculated in this paper to be significantly modified.
The reason for this is that even if the correlation length is short, the average over regions with different polarization would have a similar effect to the average over the angle between the vector field and neutrino momentum that we perform in Eq.~\eqref{angint1}.

In the scenario with a fixed overall polarization, a striking prediction of our scenario is the existence of an anisotropy in the initial dark matter velocities.
This is a consequence of the spontaneous breaking of Lorentz symmetry caused by the preferred spatial direction that the dark photon polarization singles out.
The momenta of the produced sterile neutrinos lie preferentially close to the plane orthogonal to $\vec{A}'$, meaning that the velocities in the direction of the dark photon field are initially suppressed.
For sufficiently small sterile neutrino masses for which the velocities are not completely redshifted by the beginning of structure formation, this anisotropy may be detectable in cosmological observables like the Lyman-$\alpha$ forest.

Finally, it is worth noting that for some combinations of parameters the dark photon itself can make a significant contribution to the dark matter abundance. In this case, the DM would be an admixture of dark photons and sterile neutrinos. This scenario could open additional parameter space by relaxing the constraints on the sterile neutrino mass and mixing arising from X-ray searches and structure formation observables, and perhaps even help reconcile the DM explanation of the $3.5$~keV X-ray line with recent observations that seem to challenge it~\cite{Dessert:2018qih}.
We leave this interesting possibility for future investigation.

\bigskip
{\bf Acknowledgment.} 
We thank M.~Escudero, S.~Hannestad, K.~Kainulainen, and M.~Laine for helpful correspondence, and J.~Jaeckel for thoughtful comments on the manuscript.
This work was supported by NSERC (Natural Sciences
and Engineering Research Council, Canada).
GA is supported by the McGill Space Institute through a McGill Trottier Chair Astrophysics Postdoctoral Fellowship.

\begin{appendix}
\section{Neutrino self-energy}
\label{nuse}
We can deduce the thermal contribution to the $\mu$
or $\tau$ neutrino self-energy from eqs.\ (4.48,4.52) of ref.\ \cite{Quimbay:1995jn}.  In the relativistic limit, 
the correction to the dispersion relation is parametrized as $\omega = k - b_\L$ with
\be
    b_\L = \pi\alpha_w\left\{\begin{array}{ll} {c_w^{-2}} \bar{B}(0,m_\Z) + 2\bar{B}(m_f,m_\W),& \hbox{broken phase}\\
        (2 +  {1/ c_w^2})\bar{B}(0,0).& \hbox{sym.\ phase}
        \end{array} \right.
        \label{bleq}
\ee 
Here, $c_w = \cos\theta_\W$ and the function $\bar B$ depends on the neutrino energy $\omega$ and momentum $p$, as well as $T$, $m_\Z$ and $m_f$, the mass of the charged lepton $\mu$ or $\tau$.
Treating the thermal contribution as a perturbation, we can set $\omega\cong p$, the unperturbed on-shell relation. This simplifies the form of $\bar{B}(0,m_\Z)$ to
\bea \label{eq:B0mZ}
    \bar{B}(0,m_\Z) &=& {1\over p^2}\int {\mathop{\diff k}\over 8\pi^2}\Bigg[
        \left({m_\Z^2\over 2} \frac{k}{\epsilon_z} L_2^+(k) - 4{pk^2\over \epsilon_z}\right)n_b(\epsilon_z) \nn\\
        &+&\left({m_\Z^2\over 2}L_1^+(k) - 4pk\right)n_f(k)\Bigg],
\eea
where $n_f$ and $n_b$ are the Fermi-Dirac or Bose-Einstein distribution functions, $\epsilon_z = \sqrt{k^2 + m_\Z^2}$, and
\bea
    L_2^+(k) &=& \ln\left([m_\Z^2 + 2p(k+\epsilon_z)][m_\Z^2 + 2p(k-\epsilon_z)]\over
    [m_\Z^2 - 2p(k+\epsilon_z)][m_\Z^2 - 2p(k-\epsilon_z)]\right),\nn\\
    L_1^+(p) &=& \ln\left(m_\Z^2 + 4 pk\over m_\Z^2 - 4pk\right).
\eea
This way, $B$ becomes a function of $p$ only, as we have assumed in the
text. A smooth crossover at the EWPT implies that $m_{\Z,\W}(T) \cong m_{\Z,\W} (1 - T^2/T_{EW}^2)$, which further ensures a smooth dependence on $T$ near the EWPT. 

\begin{figure}[b]
\begin{center}
 \includegraphics[width=0.99\linewidth]{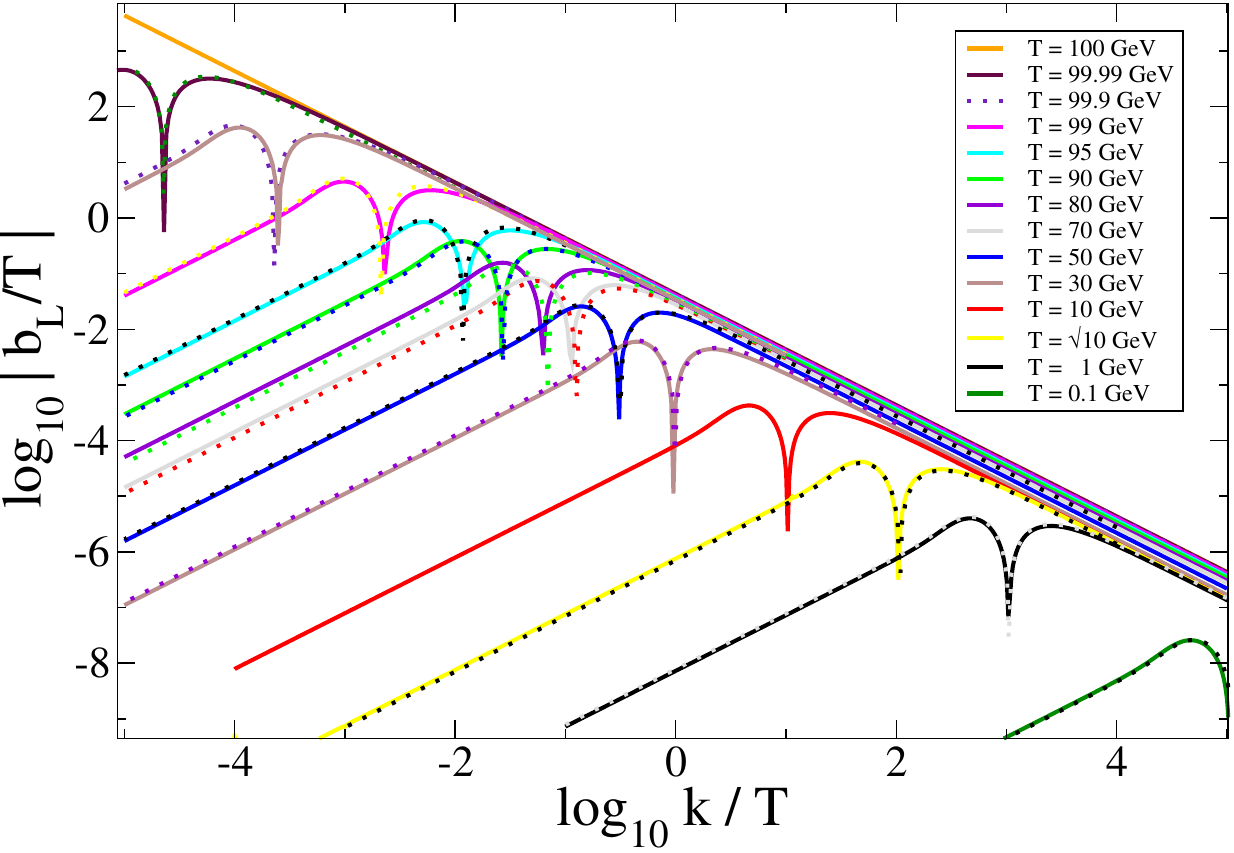}
 \caption{$|b_\L|$ versus $\mu$-neutrino momentum for a series of temperatures as shown in the legend, taking the EWPT temperature to be $T_{EW}=100\,$GeV. The dashed curves are the results of a fit as discussed in the text.}
\label{fig:bL}
\end{center} 
\end{figure}

At high temperatures, we can further simplify
$b_\L$ by taking $m_f \cong 0$ in eq.\ (\ref{bleq}), which allows us to write 
\be 
b_\L \cong \pi\alpha_w\left(2e^{-m_f/T}\bar{B}(0,m_\W(T)) + c_w^{-2}\bar{B}(0,m_\Z(T))\right)
\ee
to describe either phase, with the
factor $e^{-m_f/T}$ suppressing the
charged-current contribution for
$T < m_f$.  The neutrino thermal self-energy is thus obtained simply as $V_\W = -b_\L$.
At temperatures $T\ll T_{\rm EW}$, the integral~\eqref{eq:B0mZ} can be performed analytically, leading to
\be
V_\W (T\ll T_{\rm EW}) \simeq -\frac{14\pi}{45\alpha}G_F s_w^2 \left( 2\mathrm{e}^{-m_f/T} + c_w^2 \right)p\,T^4\,.
\ee

In Fig.\ \ref{fig:bL} we display $|b_\L|/T$ versus $p/T$ for a series of 
temperatures.  In agreement with eq.\ (\ref{eq:vweq}), $b_\L$ is negative at low $p$ and $T < T_{EW} = 100$\,GeV, while it is everywhere positive for $T> T_{EW}$.  For $T< T_{EW}$, $b_\L$ changes sign at some critical value of $p$, which decreases as
$T\to T_{EW}$ and eventually disappears. 

We find that the curves
have a nearly universal shape, such that any one of them can be obtained from another by shifting it in logarithmic space.
This saves computational effort, since then it is possible to 
obtain $|b_\L|/T$ as a function of $p/T$ at any $T$, from its functional form at a fixed
reference temperature, in terms of the two functions of $T$ that describe how the curve is shifted horizontally and vertically in the plane of
$\log_{10}|b_\L|/T$ and $\log_{10}p/T$.  These fits to the actual
values are shown as dotted curves in Fig.\ \ref{fig:bL}.

\end{appendix}

\bibliography{ref}
\bibliographystyle{utphys}

\end{document}